# Discipline Reputation Evaluation Based on PhD Exchange Network

Hua JIANG[1], Shudong YANG[1], Shuang LI[1], Shengbo LIU[1]

（1. Graduate School of Education, Dalian University of Technology, Dalian, Liaoning, China, 116024）

**Abstract:** When reputation evaluation indicators become targets, existing indicators will lose the role of indicating the true quality; At present, the evaluation of discipline reputation mostly focuses on subjective evaluation based on objective data, and there is a dispute about reliability and validity; Due to different indicators and weight settings, it is difficult to make horizontal comparison among disciplines; The evaluation also has a certain time lag. In order to solve the above four problems, this study explores a new method of discipline reputation evaluation. Taking the business administration discipline as an example, it collects data of 5848 doctoral graduates who first entered teaching posts, establishes a directed adjacency matrix from the employment unit to the doctoral degree awarding unit, and uses the theory and method of social network analysis to conduct quantitative analysis on the doctoral mutual employment network. The results show that: (1) PhD exchange network can explain discipline reputation and is a new indicator to measure discipline reputation; (2) From the perspective of employment behavior among colleges and universities, there is horizontal flow and downward flow between the head colleges and universities, and downward flow is mainly among the middle and lower colleges. There is a time lag between college talent recruitment and academic achievement output. Therefore, the mining of the structural characteristics and network evolution trend of the PhD exchange network based on the "foot voting" of doctoral graduates is faster than the discipline ranking based on the follow-up achievement indicators to reflect the changes in the discipline quality, which can be used to warn the changes in the discipline quality.

## 1. Introduction

There are many uncertain factors in the evaluation of higher education, one of which is the lack of solid and objective data to measure the reputation level of a university. Such biased evaluation results will react on the process of university goal setting and talent training, thus leading to the alienation of university competition[1]. According to Goodhart's Law, when reputation evaluation indicators become the established goals of colleges and universities to guide the formulation of teaching policies, the evaluation indicators will lose their original information value[2], because the teaching policy makers will sacrifice other aspects to strengthen this indicator, thus making the existing indicators lose the role of indicating the true reputation level.

At present, the mainstream evaluation method of discipline reputation is reputation evaluation based on index system and weight. However, unlike scientific research evaluation that is relatively easy to quantify, the evaluation of discipline reputation is mostly subjective evaluation based on objective data, resulting in new deviations. These biased results will directly affect the knowledge production mode, the behavior, values and sociological imagination of stakeholders such as teachers and students[3]. As an abstract concept, discipline reputation is difficult to quantify, and as a subjective experience, it is difficult to reach consensus[4]. Therefore, new consensus measurement models and supporting calculation methods are needed to explore the evaluation methods of discipline reputation from a new perspective.

With the continuous expansion of China's doctoral training scale and the gradual relative saturation of the academic labor market, the doctoral degree has become the entry threshold for most colleges and universities. Although doctorate employment shows a trend of diversification, teaching posts are still the first choice for doctorate employment[5]. In recent years, 64.23% of the doctoral graduates of Tsinghua University have obtained academic posts[6], and 59.32% of the dual first-class universities[7]. Through systematic academic training, doctoral students have strengthened their discipline loyalty[8], which provides a basic guarantee for the closeness of the "doctor faculty" relationship network[9].

There are many stakeholders in the academic labor market, and their relationships are complex. The expectations of stakeholders on the academic labor market also constitute one of the realistic motivations for the development of disciplines[10]. There is a certain contradiction between job seekers and employment decision-makers in the academic labor market. On the one hand, job seekers give priority to the employment unit with the highest reputation within their ability, and generally use the "high medium low" job search strategy to realize their interest demands; On the other hand, employment decision-makers generally require job seekers to graduate from famous schools, at least not lower than their schools, and generally realize their interest demands through the formulation of personnel policies, the provision of interview opportunities, interview scoring, interview voting and other means. The intersection between the applicant and the employer is a necessary condition for successful application, and the existence of this intersection also reflects the stability of the discipline point's reputation rank order.

In a word, the four characteristics of doctoral degree access threshold, more than half of academic reproduction, discipline loyalty, and stability of reputation rank order have facilitated the research related to the "doctoral academic labor market". The emerging method of evaluating discipline reputation based on the doctoral academic labor market has gradually become a research hotspot. This new evaluation method is an effective supplement to the traditional evaluation method based on input and process data[11], It is helpful to make up for the shortcomings of existing ranking and evaluation.

# 2. Related research

## 2.1. Discipline reputation evaluation based on index system

The evaluation method of discipline reputation based on indicator system is based on reputation metrics and reputation survey. The premise of implementing the reputation indicator method is to build a complete multi-level reputation evaluation indicator system and set weights[12]. The representative case is the Boston Reputation Research Institute led by Fombrun, an American reputation management scholar, which conducts a reputation quotient evaluation based on the data provided by Harris Interactive Market Research Company. In the field of academic reputation evaluation, common applications are based on expert interviews, questionnaires, structural equation models, etc. to determine the factors and weights that affect discipline reputation[13]. The reputation evaluation in the mainstream world discipline ranking generally adopts the reputation index method, and the comparison is shown in the following table:

**Table 1 Comparison of reputation evaluation in the ranking of business management disciplines**

| | GRAS: business administration | THE: China Subject Ratings: Business Administration | QS World University Rankings by Subject: Business & Management Studies | USNews SUBJECT RANKINGS: Economics and Business |
|---|---|---|---|---|
| **Proportion of reputation survey** | 0.0% | 56.6%，including 28.3% teaching reputation and 28.3% Scientific reputation | 50.0%, including 40.0% academic reputation and 10.0% employer reputation | Scientific reputation: 25.0% |
| **Investigation cycle** | / | First quarter of each year | each year | each year |
| **University coverage** | 400 universities | 795 universities | 573 universities | 250 universities |
| **Respondents** | / | Elsevier Scholars Library | Own scholar library | Own scholar library |
| **sampling method** | / | Stratified random sampling | Stratified random sampling | Stratified random sampling |
| **Sample size** | / | / | Academic reputation: 130000<br>Employer reputation: 75000 | / |
| **Evaluation method** | / | Reputation index method | Reputation index method | Reputation index method |
| **Lobbying agencies** | / | The World 100 Reputation Network | / | / |
| **Reputation scoring rules** | / | Based on the number of institutional nominations, weighted by country and discipline, and standardized by | / | / |

| | | logarithmic function | | |
|---|---|---|---|---|

It can be seen from the above table that, from the perspective of time cycle, the data collection cycle of discipline ranking based on the indicator system is long, which is generally conducted once a year, and the evaluation results have a certain time lag effect. From the perspective of data objectivity, in the United States where the lobbying system is legalized, the manipulation of reputation by lobbyists will affect the objectivity of reputation data. In addition, in the actual operation of indicator system design and weight distribution, it is actually a subjective evaluation based on objective data, resulting in weak credibility. The data relied on by the reputation evaluation method based on the indicator system are not only behavioral data, but also survey data. The survey data may be unintentionally or intentionally deceptive, or lead to survivorship bias and other data errors due to improper sampling methods, which also lead to weak credibility of the reputation evaluation method based on the indicator system.

As a special category of education evaluation extending to the field of higher education, the evaluation elements, weights and algorithms of discipline reputation evaluation are diversified due to the multiple needs of stakeholders. When the needs of all parties cannot be met at the same time, there will be evaluation difficulties, such as invalid indicators, difficult consensus, cascading evaluation bias, difficult horizontal and vertical comparison, evaluation delay, etc.

1) **Invalid indicators**. According to Goodhart's Law, when discipline evaluation indicators become the established goals of universities to guide policy formulation, the evaluation indicators will lose their original information value, because university policy makers will sacrifice other aspects to strengthen the indicators, so that the existing reputation evaluation indicators lose the role of indicating the real reputation[14][15].

2) **Difficult consensus**. At present, the mainstream evaluation method for discipline reputation is based on index system and weight, which is different from scientific research evaluation that is relatively easy to quantify. The evaluation of discipline reputation is mainly subjective evaluation based on objective data, which cannot guarantee the objectivity of evaluation method and index weight, the integrity and consistency of evaluation data[16]. In addition, There is a certain gap between the credibility level of third-party education assessment institutions in China and that of foreign education assessment institutions[17].

3) **Cascading evaluation bias**. Discipline evaluation is also affected by globalization, national administrative intervention, and bibliometric indicators[18], resulting in new deviation[4]. These biased results will directly affect the mode of knowledge production, the behavior of stakeholders[19], values and sociological imagination[3].

4) **Difficult horizontal and vertical comparison**. On the one hand, due to the differences of disciplines, the reputation evaluation system generally sets different indicator structures and weights for different disciplines[7], which makes it difficult to make horizontal comparisons between disciplines. On the other hand, the reputation evaluation system of the same discipline will also change over time, which makes it difficult to make vertical comparison on time series.

5) **Evaluation delay**. Due to the time lag effect of education input and output[20], this has led to the lag of discipline reputation evaluation. For example, the evaluation of talent training quality in China's discipline evaluation index system is based on high-level alumni, while there is a time lag of several years from graduation to a certain social status.

## 2.2. Discipline Reputation Evaluation Based on Reputation Network

With the progress of information technology and network science, reputation network theory provides a new opportunity to solve the above dilemma. This theory believes that reputation is a consensus established from social networks[21]. Reputation networks use objective criteria to track the contribution of each actor to the network, so we can use the numerical calculation of relevant social networks to measure reputation consensus[22]. The so-called objective standards are carriers that are easy to be collected and recognized, such as innovation relationship networks and citation networks. For example, the EU nanotechnology innovation network is built based on the cooperation between academic organizations, and the academic reputation is quantified by node centrality[23]. Citation network based on academic literature citation behavior, and then use PageRank algorithm to conduct social network econometric analysis, and then evaluate the reputation of academic institutions[24]. Based on the mobility behavior of students in the three stages of undergraduate, postgraduate and doctoral studies, we can build a mobility network of universities, and then measure the reputation of universities[25].

The evaluation method based on reputation network theory needs to use a certain social network as the carrier for social network numerical calculation, and the establishment of these social networks is based on the collection of actual social behavior and subsequent results. In theory, contemporary colleges and universities have begun to play the role of assigning grades on the career ladder and positions in the social structure[26]. In practice, colleges and universities usually tend to pursue the accumulation and reproduction of talent training quality by recruiting first-class scholars and attracting excellent students[27]. Therefore, one of the entry points to evaluate the discipline reputation is the employment behavior of doctoral graduates.

## 2.3. The model of PhD exchange network

The recruitment of teaching posts by university doctoral program can be seen as the implicit positive evaluation of the recruitment unit on the training unit[28], which is also the logical starting point for evaluating the discipline reputation based on the doctoral recruitment data. The introduction of doctoral graduates by colleges and universities is a social exchange network that connects a national or even a global discipline institution and forms a relationship between doctoral mutual employment. The two-way choice between doctoral graduates and universities is a rational interactive choice based on the comprehensive consideration of maximizing their own interests, which naturally emerges to form peer review among university doctoral program. Therefore, this social exchange network is objective, stable, and easy to collect data[29].

The social capital of academic institutions is a function of the number and mode of connections between academic institutions established through social exchanges between academic institutions[30], that is, the social capital of each academic institution can be calculated by measuring the social exchanges between academic institutions, such as talent exchanges and literature citations. One of the formalized carriers of social exchange is the PhD exchange network based on talent employment behavior. The PhD exchange network is a social measurement index of the social capital of academic institutions. In turn, social capital is a function of the PhD exchange network[31].

PhD exchange network is an employment relationship network formed by academic institutions to employ doctoral graduates in the academic labor market. Based on the eigenvector centrality algorithm, the social capital in the PhD exchange network is calculated, and then the reputation of academic institutions is indirectly measured. It is found that 82.1% of the difference in academic institutions' reputation can be explained by a single variable of calculated social capital through regression analysis[31]. From the perspective of the "center edge" hierarchical structure, academic institutions in different hierarchical positions in the PhD exchange network have different strengths in talent attraction and academic resource control, which reflects the size of social capital of academic institutions[32].

The research of domestic and foreign scholars for many years has enriched the connotation of the PhD exchange network, which is mainly reflected in the expansion of talent flow direction, network computing algorithm, academic inbreeding, time series analysis, etc.

Schmidt et al. first used the directed network in the PhD exchange network and introduced the PageRank algorithm, an Internet page ranking algorithm, to perform numerical calculation[33]. In reality, there is academic inbreeding in doctoral recruitment. Academic inbreeding behavior corresponds to the concept of "self loop" in network science. The introduction of the concept of self loop can faithfully reflect the real employment structure[33]. Behind every concrete complex system, there is an abstract complex network to depict the interaction between various elements in the complex system[34]. Therefore, the PhD exchange network is essentially an abstract embodiment of the social exchange system, reflecting the static structure and evolution law of the academic labor market.

In practice, discipline reputation is hierarchical. The reproduction of reputation hierarchy depends on two points: first, privileged members adopt unique working and living styles, such as academic forums and works; Second, close the social exchange between the upper and lower levels, such as the closed academic circle and academic titles[35]. Burris further conceptualizes reputation on the basis of the above views: it is owned by individual institutions, reproduced by closing the social exchange between high and low status groups, and can be measured by the size of social capital. Reputation is the most important form of social capital of academic institutions. It is found that its interpretation of academic labor market exchange relationship is far more than academic productivity[31]. The stratification of reputation leads to the monopoly of opportunities[35], which is also true in the academic field: the appointment of doctoral graduates between academic institutions can be regarded as the affirmation of the reputation of the other party's talent training quality, a symbolic act of mutual endorsement, and in this process, the social exchange of academic talents has been completed, while the PhD exchange network is an abstract and real-time update of this social exchange. The accumulation process of academic institutions' reputation is operated through the self reproduction capacity of social capital, while the PhD exchange network is an abstract description of the social exchange of reputation, which is also a social measurement index of academic institutions' social capital.

First, observe the single employment behavior from the micro level. The first step is that the doctoral training unit will conduct personnel training for its doctoral postgraduates. When the students meet the requirements of talent training quality objectives, they will be awarded the doctoral degree. The second step is that doctoral graduates become the carrier of the quality signal of talent training in the academic labor market, and the employer's examination of the student is equivalent to the identification of the quality signal. Third, the employer's behavior of employing the student is equivalent to a reputation vote on the training unit.

Secondly, the employment behavior between the two academic institutions is observed from the meso level. Iterating the single employment behavior, we can find two employment modes: 1) **One way employment mode**, that is, the two academic institutions are a one-way relationship of "training unit - employer"; 2) **Exchange employment mode**, that is, the two academic institutions are "training units - employers".

Finally, the employment behavior of all academic institutions is observed from the macro level. Integrate the employment behavior between all two academic institutions to form a PhD exchange network. The network node represents the academic institution, and the edge of the network represents the employment behavior. The direction of the edge is from the employer to the training unit, that is, the direction of reputation voting. The conceptual model of discipline reputation evaluation is shown in the figure below:

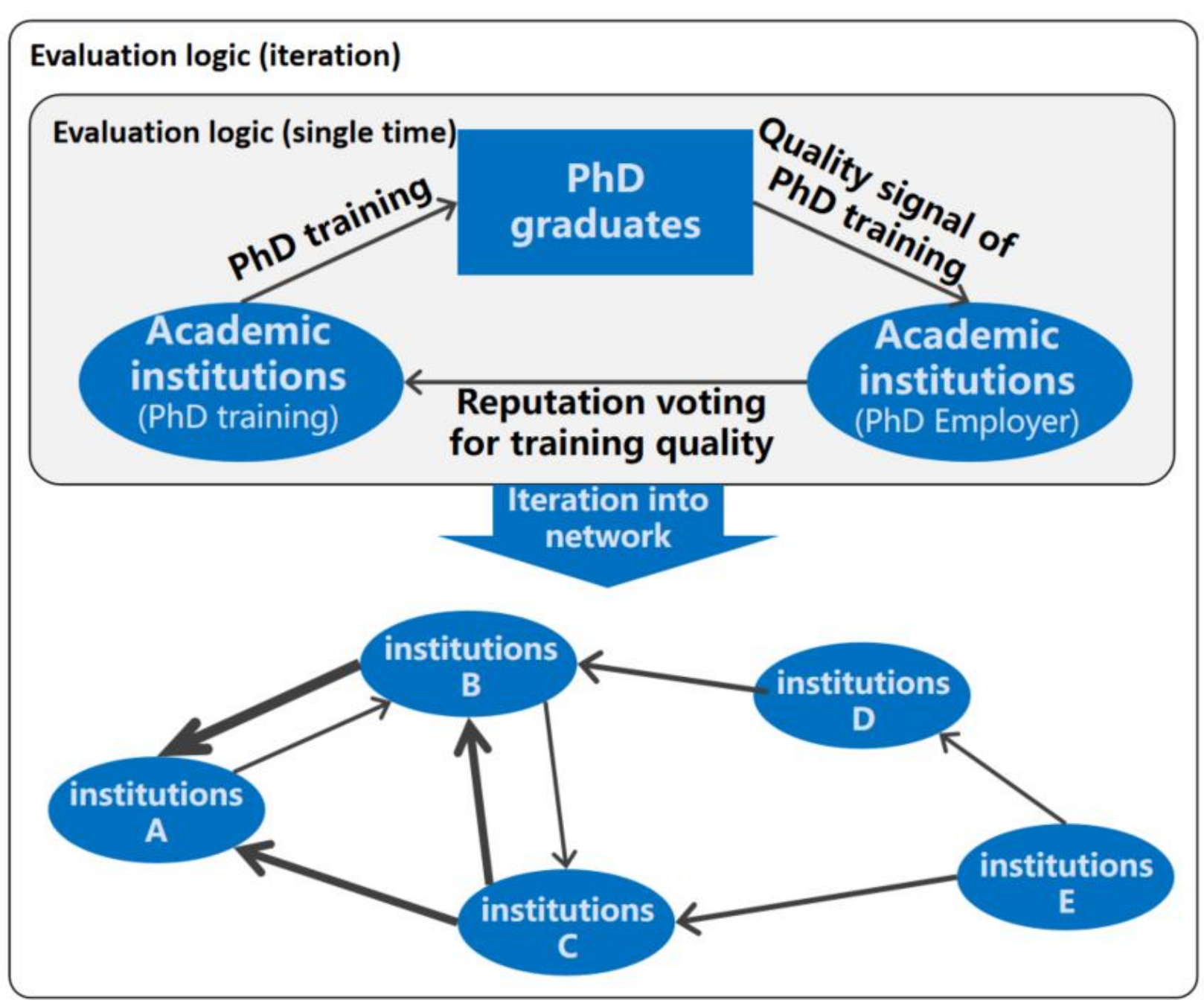

**Figure 1 Conceptual Model of Discipline Reputation Evaluation**

The social exchange in this conceptual model is mainly reflected in the rational choice between academic institutions to hire doctoral graduates. The social econometric indicators contained in the PhD exchange network have rich connotations, such as the talent scale, talent flow, time series, etc. of the doctoral program, which can accurately depict the talent training quality of academic institutions.

Observing the academic labor market in China, there are problems such as information asymmetry. Driven by the minimization of identification costs and risks, employers often select talents according to the "famous school complex"[36] or "educational lineage"[37], that is, the human resource allocation of doctoral graduates in the academic labor market follows the "downward flow mode"[38], and this directed flow mode is not affected by supply and demand[39]. Due to the differences in the structure, scale and characteristics of academic labor markets at home and abroad, the research methods and conclusions on the PhD exchange network abroad may not be applicable to China. However, the sample size of doctoral employment and the coverage of universities are relatively limited in domestic related studies. Therefore, based on the above existing research results and dilemma analysis, it can be concluded that the research question is whether the discipline reputation can be evaluated through the directed multi-layer network of mutual recruitment of PhD graduates under the directed flow mode of PhD graduates?

## 3. Research Design

As a social entity, colleges and universities are actors in the social network. The position of actors in the social network and the relationship structure between actors have an important impact on the behavior of actors. For the research of social relationship structure, social network analysis method is a research paradigm with mature technology and wide application up to now[32]. Compared with traditional structural sociology, social network analysis based on the relationship perspective is a powerful tool to re understand the nature, strength, scale, stratification and other structural characteristics of the relationship between actors, and can reveal the ability of actors to absorb social resources embedded in social networks.

In order to avoid the deviation introduced by the differences between disciplines, a single discipline, business administration, which covers a wide area and involves a large number of universities, is selected as the research object. Collect relevant employment data from 172 doctoral training universities and 240 doctoral employment universities under the first level discipline of business administration, and 5980 doctoral graduates were recruited into the teaching posts of business administration discipline for the first time. The data collection channel is mainly based on the teacher homepage and scholar database of the official website of the doctoral graduates' employment colleges. If the data is incomplete, the data will be supplemented with reference to relevant news, announcements, school yearbooks, academic resumes, etc. The data will be cross verified according to the information of doctoral dissertations, and 5848 valid samples will be finally obtained.

After data collection, it is stored in the relational database. Important fields include name, degree awarding unit, employment unit, graduation year, employment year, etc. If and only if the year of employment is later than the year of graduation, it will be regarded as the first time for doctoral graduates to enter the teaching position. For the subsequent social network analysis, Python programming language is used to automatically convert relational data into a directed adjacency matrix, with the direction pointing to the source of talents - degree awarding units. In order to reflect the real network structure, the self employment data is retained, and the data on the diagonal of the adjacency matrix is not set to zero. Finally, the 240 order (1 node for all overseas doctoral awarding units and 239 domestic universities) weighted adjacency matrix that can truly reflect the mutual employment of doctors is obtained.

Centrality is one of the indicators that describe the position of actors in the social network and the strength of their relationship with other actors. Common centrality includes degree centrality, proximity centrality, intermediate centrality, eigenvector centrality, etc. Bonacich's definition of eigenvector centrality is that the importance of a node in a social network depends on both the number of its neighbors and the importance of its neighbors[40]. Bourdieu's definition of social capital based on the theory of relational network resources is: the actual or potential resources brought to people by their position in the persistent network structure, including two dimensions of quantity and quality[30]. Therefore, Bonacich's definition of eigenvector centrality is consistent with Bourdieu's definition of social capital in terms of operation. The stock of social capital can be represented by eigenvector centrality. In the PhD exchange network, the reputation of talent training quality of each university can be regarded as a kind of social capital. Therefore, this study uses the eigenvector centrality to measure the reputation of each university discipline point emerging in the game of mutual recruitment of doctors, so as to reflect the reputation level of the discipline point.

In order to improve the reusability and flexibility, this research adopts the method of self programming for social network analysis. Based on Python programming language, NetworkX social network analysis software package, Pandas data analysis software package, Matplotlib data visualization software package, etc., the directed adjacency matrix is calculated graphically to obtain the eigenvector centrality and visual network diagram, which are used to visualize the structural characteristics of the PhD exchange network.

# 4. Research findings

## 4.1. Morphological change trend of the PhD exchange network

Through social network computing and network visualization, the PhD exchange network has a total of 240 university discipline nodes, 1942 directed edges, and the average in and out degree is 8.09. The network visualization is shown in Figure 2, including five sub graphs. Sub graph a is the PhD exchange network from 1980 to 2000, sub graph b is the PhD exchange network from 2001 to 2007, and sub graph c is the PhD exchange network from 2008 to 2014, Sub graph d is the PhD exchange network from 2015 to 2021, and sub graph e is the PhD exchange network in the whole time range.

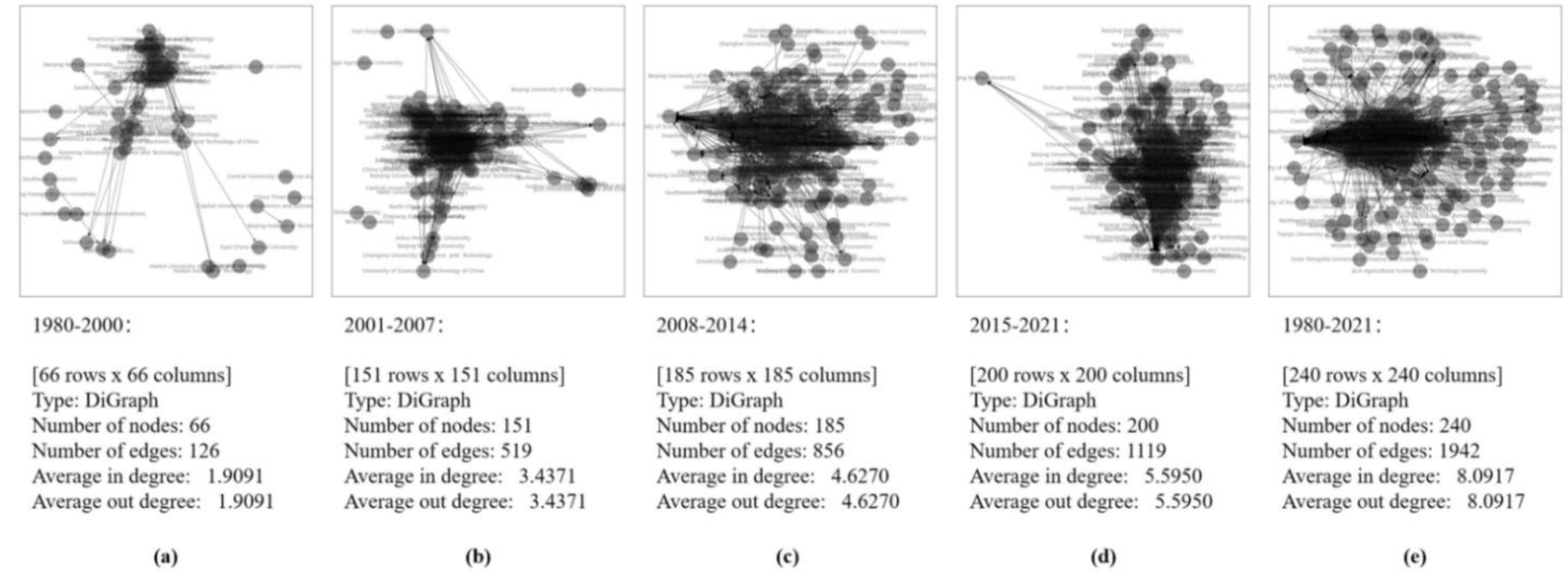

**Figure 2 Visualization of PhD exchange network**

From the overall perspective, it can be seen from the above figure that the doctoral candidates employed by university doctoral program mainly come from "985" universities and overseas universities. In terms of the data collected so far, only a few graduates from domestic universities have successfully worked in overseas universities, such as Tsinghua University. This shows that the PhD exchange network generated by domestic universities and the PhD exchange network generated by overseas universities are basically in a "one-way interaction" state. These few universities play a "structural hole" role in connecting the foreign and domestic PhD exchange networks. It also shows that the domestic and overseas PhD exchange networks belong to two network communities, The conclusion of overseas PhD exchange network may not be applicable to domestic.

From the perspective of time series, prior to 2008, there were isolated academic nodes in China's PhD exchange network (sub graph a and sub graph b of Figure 2). The existence of these isolated nodes indicates that these universities were mainly self employed or mutually employed. Since 2008, China's PhD exchange network has gradually become "closed" (sub figure c of Figure 2), forming a closed relationship of mutual employment between universities and gradually breaking the academic "inbreeding". Over time, the number of doctoral network nodes, the number of network directed edges, the average degree of entry and the average degree of exit are all increasing, which reflects the increasing number of doctoral program and the increasing mobility of talents across regions. From the perspective of employment behavior among colleges and universities, there is horizontal flow and downward flow between the head colleges and universities, and downward flow is mainly among the middle and lower colleges.

## 4.2. Analysis of the Influence of Academic Relationship Structure Preference on Discipline Reputation

The self employment ratio, the overseas ratio, the ratio of newcomers in recent five years, the scale of PhD, and Tsinghua-Peking ratio are taken as independent variables, and the eigenvector centrality value of the PhD exchange network in all time ranges is taken as dependent variables for linear regression analysis. The R-squared value of the model is 0.762, which means that the above five independent variables can explain the change of 76.2% of the dependent variables. When conducting F test on the model, we found that the model passed the F test (F=31.294, p=0.000<0.05), which means that at least one of the above five independent variables will have an impact on the dependent variable. In addition, we found that the VIF values in the model are all less than 5, which means there is no colinearity problem; The D-W value is close to the number 2, which indicates that the model does not have autocorrelation, and there is no correlation between sample data. The model has a good effect. The regression analysis results are shown in the following table:

**Table 2 Results of linear regression analysis (*n*=55)**

| | Denormalization coefficient | | Standardization coefficient | *t* | *p* | VIF | $R^2$ | Adjusted $R^2$ | *F* |
|---|---|---|---|---|---|---|---|---|---|
| | *B* | Standard error | *Beta* | | | | | | |

**Table 2 Results of linear regression analysis (*n*=55)**

| | Denormalization coefficient | | Standardization coefficient | *t* | *p* | VIF | $R^2$ | Adjusted $R^2$ | *F* |
|---|---|---|---|---|---|---|---|---|---|
| | *B* | Standard error | *Beta* | | | | | | |
| constant | -0.130 | 0.069 | - | -1.887 | 0.065 | - | 0.762 | 0.737 | *F* (5,49)=31.294,*p*=0.000 |
| self_ratio | 0.383 | 0.092 | 0.309 | 4.147 | 0.000** | 1.142 | | | |
| overseas_ratio | 1.143 | 0.160 | 0.671 | 7.158 | 0.000** | 1.804 | | | |
| new_ratio | -0.250 | 0.176 | -0.107 | -1.421 | 0.162 | 1.162 | | | |
| n | 0.001 | 0.001 | 0.123 | 1.431 | 0.159 | 1.520 | | | |
| TsPek_ratio | 0.516 | 0.311 | 0.145 | 1.661 | 0.103 | 1.558 | | | |

* $p<0.05$ ** $p<0.01$

There is a significant positive relationship between the overseas ratio and discipline reputation, that is, the overseas ratio of doctoral program with higher discipline reputation is also higher. From the perspective of school statistics, there is also a significant positive relationship between self employment ratio and discipline reputation, that is, the self employment ratio of doctoral program with higher discipline reputation is also higher. However, if schools are classified according to their levels, the self employment ratio presents an inverted U-shaped curve as the school level decreases. the ratio of newcomers in recent five years, the scale of PhD, and Tsinghua-Peking ratio have no significant impact on the discipline reputation.

## 4.3. Analysis of the Influence of Academic Relationship Structure Preference on the Trend of Discipline Reputation

The self employment ratio, the overseas ratio, the ratio of newcomers in recent five years, the scale of PhD, and Tsinghua-Peking ratio are taken as independent variables, while the moving average obtained from the time slicing of the PhD exchange network represents the trend of reputation change, and is taken as a dependent variable for linear regression analysis. It does not pass the F test (F=0.815, p=0.545>0.05), which means that the above five independent variables will not affect the trend of discipline reputation.

**Table 3 Results of linear regression analysis (*n*=55)**

| | Denormalization coefficient | | Standardization coefficient | *t* | *p* | VIF | $R^2$ | Adjusted $R^2$ | *F* |
|---|---|---|---|---|---|---|---|---|---|
| | *B* | Standard error | *Beta* | | | | | | |
| constant | 0.943 | 0.053 | - | 17.830 | 0.000** | - | | | |
| self_ratio | 0.001 | 0.071 | 0.001 | 0.010 | 0.992 | 1.142 | 0.077 | -0.017 | *F* (5,49)=0.815,*p*=0.545 |
| overseas_ratio | -0.068 | 0.122 | -0.102 | -0.552 | 0.584 | 1.804 | | | |

**Table 3 Results of linear regression analysis (*n*=55)**

| | Denormalization coefficient | | Standardization coefficient | $t$ | $p$ | VIF | $R^2$ | Adjusted $R^2$ | $F$ |
|---|---|---|---|---|---|---|---|---|---|
| | $B$ | Standard error | *Beta* | | | | | | |
| new_ratio | 0.056 | 0.135 | 0.062 | 0.418 | 0.678 | 1.162 | | | |
| n | -0.000 | 0.001 | -0.042 | -0.247 | 0.806 | 1.520 | | | |
| TsPek_ratio | 0.397 | 0.238 | 0.285 | 1.663 | 0.103 | 1.558 | | | |

* $p<0.05$ ** $p<0.01$

The preference of academic relationship structure has no significant influence on the trend of discipline reputation. The short-term operation strategy that the doctoral program try to improve the discipline reputation by adjusting the academic relationship structure is not significant in reality.

# 5. Method validation

If the eigenvector centrality of university discipline points in the PhD exchange network is related to the actual score in the discipline ranking, it verifies that the position of university doctoral program in the PhD exchange network can explain the level of discipline reputation. Therefore, the following research assumptions are drawn:

***H1***: The feature vector centrality of university doctoral program in the PhD exchange network is positively related to the discipline ranking score of the university

***H2***: The eigenvector centrality of university doctoral program in the PhD exchange network is positively correlated with the rank in the discipline evaluation of the university

**Table 4 Validation Data (Top 10 universities)**

| # | Universities | EC-2007 | EC-2014 | EC-2021 | GRAS 2021 | Rank of the 3rd round of discipline evaluation | Rank of the 4th round of discipline evaluation |
|---|---|---|---|---|---|---|---|
| 1 | Tsinghua University | 1.0000 | 0.9403 | 1.0000 | 795 | 15 | 9 |
| 2 | Peking University | 0.4967 | 0.6309 | 0.6656 | 799 | 13 | 8 |
| 3 | Renmin University of China | 0.3952 | 0.5585 | 0.3993 | 859 | 16 | 9 |
| 4 | Xi'an Jiaotong University | 0.5930 | 0.6408 | 0.4103 | 571 | 16 | 8 |
| 5 | Shanghai Jiaotong University | 0.8269 | 0.6712 | 0.4098 | 718 | 12 | 9 |
| 6 | Nankai University | 0.6208 | 1.0000 | 0.6815 | 625 | 13 | 8 |
| 7 | Fudan University | 0.7308 | 0.6861 | 0.5477 | 398 | 11 | 8 |
| 8 | Xiamen University | 0.6710 | 0.5944 | 0.2906 | 788 | 14 | 8 |
| 9 | Huazhong University of Science and Technology | 0.4945 | 0.5638 | 0.3887 | 431 | 9 | 7 |

| 10 | Tianjin University | 0.9267 | 0.9325 | 0.4744 | 287 | 10 | 6 |
|---|---|---|---|---|---|---|---|

Note: Due to the lack of public data, the University of International Business and Economics, Shandong University, Central University of Finance and Economics and other universities are not in the list, so this study does not represent the true ranking

Because the results of the third round of discipline evaluation and the fourth round of discipline evaluation are inconsistent, they are uniformly converted into "rank" numerical variables, such as "C -" is counted as "1", "C" is counted as "2", "A+" is counted as 9, and "0" when the discipline point does not participate in the discipline evaluation. After transformation, all the variables in the above table are numerical variables, so Pearson correlation analysis can be used to study the correlation between the centrality of the feature vector of the PhD exchange network, the score of GRAS 2021, and the rank of discipline evaluation. Pearson correlation coefficient is used to express the strength of the correlation. The calculation results are shown in Table 4:

**Table 5 Pearson correlation coefficient (n=229)**

| | **Rank of the 3rd round of discipline evaluation** | **Rank of the 4th round of discipline evaluation** | **GRAS 2021** |
|---|---|---|---|
| EC-2014 | **0.6491**** | / | / |
| EC-2021 | / | **0.6157**** | **0.7424**** |

* p <0.05 ** p <0.01

Before calculating Pearson correlation coefficient, 10 universities with incomplete data among 239 domestic universities were eliminated, and the total number of university samples was 229. According to the EC-2014 value corresponding to the third round of discipline evaluation results, EC-2021 value corresponding to the fourth round of discipline evaluation results, and EC-2021 value corresponding to the GRAS 2021 score, Pearson correlation coefficient in the above table is obtained. The correlation coefficient between EC-20014 value and the third round of discipline evaluation rank is 0.6491, showing a significant level of 0.01. The correlation coefficient between EC-2021 value and the fourth round of discipline evaluation rank is 0.6157, showing a significant level of 0.01. Therefore, there is a significant positive correlation between the centrality of the feature vector of the PhD exchange network and the discipline evaluation rank; The correlation coefficient between the EC-2021 value and the GRAS 2021 ranking score is 0.7424 and presents a significance level of 0.01, which indicates that there is a significant positive correlation between the centrality of the feature vector of the PhD exchange network and the GRAS 2021 ranking score, so the H1 and H2 assumptions are valid.

# 6. Conclusion and discussion

This research takes the centrality of the eigenvector representing social capital in the network as the ranking algorithm, and takes the first employment behavior of teachers in the first level discipline of business administration in China as the sample, trying to explore how to evaluate the discipline reputation based on the PhD exchange network under the "downward flow mode" of doctoral graduates.

In essence, the PhD exchange network is an abstract embodiment of the social exchange system in the academic labor market, which can reflect the static structure of the academic labor market. The discipline reputation of each university in the PhD exchange network is regarded as a kind of social capital, and the stock of this social capital is characterized by the centrality of the eigenvector to quantitatively reflect the discipline reputation of the doctoral training unit. Through numerical calculation, the research found that the eigenvector centrality of the PhD exchange network was positively correlated with the rank of discipline evaluation and GRAS ranking score, and both showed a significant level of 0.01. From the perspective of employment behavior among colleges and universities, there are horizontal and downward flows between the head colleges and universities, and downward flows between the middle and lower colleges and universities, which is also a confirmation of the labor market segmentation theory.

The above verifies that the structure and numerical characteristics of the PhD exchange network are a consensus network that can reflect the reputation of disciplines among universities. Compared with the reputation evaluation method based on the indicator system, the principle of this network method is simpler, the data is easier to collect and cross verify, and the operation does not contain subjective factors, so it has higher reliability, validity and operability. In addition, the emerging method of evaluating discipline reputation based on the output performance of the labor market avoids the shortcomings of the evaluation method based on input and process data, such as the imbalance of teaching and research, the lack of timeliness, and the inability to make horizontal comparisons between disciplines and vertical comparisons in time due to excessive reliance on bibliometric data.

Compared with the discipline reputation evaluation method based on the indicator system design, the discipline reputation evaluation method based on the doctorate mutual employment behavior can reflect the discipline reputation more objectively, stably, rapidly, scaled, low-cost and comprehensively. 1) Objectivity is reflected in the fact that all research data are based on real application behavior; 2) Stability is reflected in the simple rule that the recruitment of doctoral candidates in discipline sites is regarded as reputation voting, which can be compared vertically across time, and even the reputation of disciplines without discipline evaluation and ranking in the last century can be studied; However, the complex rules based on indicator system and weight lose the possibility of longitudinal time comparison or introduce deviation once the indicators or weights are changed; 3) The rapidity is reflected in the fast updating cycle of online teacher information of discipline point officials; 4) The scalability is reflected in the micro analysis of the reputation of doctoral program, and the macro analysis of the discipline development level of schools in different regions and at different levels after classification; 5) The economy is reflected in that the automatic method can be used to obtain the research data, and the collection cost is lower than the questionnaire survey. 6) Comprehensiveness is reflected in the ability to comprehensively observe the academic relationship structure of doctoral program and infer the talent preference of doctoral program, including but not limited to: self employment, overseas employment, intra city employment, employment of Tsinghua Peking University graduates, employment in recent five years, etc.

Compared with other similar researches, the previous research models mainly focus on the undirected network without self loop. The innovation of this research is that the time series analysis of the directed network with self loop is realized through self programming. Self loop can reflect the self employment behavior of colleges and universities, directed network can reflect the flow direction between colleges and universities, and time series analysis can reflect the evolution trend of academic labor market. This network model can fit the academic labor market more realistically, which is a potential supplement to existing related research.